\documentclass[journal,comsoc]{IEEEtran}
\usepackage[T1]{fontenc}

\newcommand{\argmin}{\mathop{\rm arg~min}\limits}

\usepackage{graphicx}
\ifCLASSINFOpdf
\else
\fi
\usepackage{amsmath}
\usepackage[linesnumbered,ruled,vlined]{algorithm2e}
\usepackage{algorithmic}
\usepackage[cmintegrals]{newtxmath}
\begin{document}
%
\title{Randomization Approaches for Reducing PAPR with Partial Transmit Sequences and Semidefinite Relaxation}
%
%
%

\author{Hirofumi~Tsuda,~\IEEEmembership{Student Member,~IEEE,}
        Ken~Umeno,~\IEEEmembership{Member,~IEEE}}

%
%

\markboth{Journal of \LaTeX\ Class Files,~Vol.~14, No.~8, August~2015}%
{Shell \MakeLowercase{\textit{et al.}}: Bare Demo of IEEEtran.cls for IEEE Communications Society Journals}
%



\maketitle

\begin{abstract}
To reduce peak-to-average power ratio, we propose a method to choose a suitable vector for a partial transmit sequence technique. With a conventional method for this technique, we have to choose a suitable vector from a large amount of candidates. By contrast, our method does not include such a selecting procedure, and consists of generating random vectors from the Gaussian distribution whose covariance matrix is a solution of a relaxed problem. The suitable vector is chosen from the random vectors. This yields lower peak-to-average power ratio, compared to a conventional method for the fixed number of random vectors.
\end{abstract}


%
\IEEEpeerreviewmaketitle

\section{Introduction}
%
%
%
%
\IEEEPARstart{O}{rthogonal} Frequency Division Multiplexing (OFDM) systems are widely used and their signals are generated by Inverse Fast Fourier Transformation (IFFT) \cite{ofdmcdma}. One advantage for OFDM systems is that OFDM systems can deal with multi-path delay. Since OFDM systems are implemented by IFFT, effects of multi-path delay in frequency selective channels can be removed with guard interval techniques and zero padding techniques \cite{zp}. Due to the resistance to flat fading channels, Multiple-Input Multiple-Output systems with OFDM systems have been investigated \cite{mimo}.

While there are some advantages in OFDM systems, there are two main problems. One is that signals of OFDM systems have relatively large side-lobes \cite{filterbank}. This problem is caused since OFDM signals consists of sine waves. The other is that signals of OFDM systems have large Peak-to-Average Power Ratio (PAPR), which is the ratio of the maximum value of RF signal powers to the average value of them. Approximately, the output power grows linearly for low values of the input powers. However, for input signals with large power, the growth of the output power is not linear. Then, in-band distortion and out-of-band distortion are caused for a large input power \cite{concept}. With symbols chosen independently, PAPR for OFDM signals has been investigated in \cite{ochiai} \cite{extreme}. With dependent symbols, for example, Bose-Chaudhuri-Hocquenghem (BCH) codes, their PAPR has been investigated \cite{general}. Further, the performance of OFDM systems with non-linear amplifiers has been investigated in \cite{clip} \cite{costa}.

To reduce PAPR, many methods have been proposed and explored. For example, a selected mapping method \cite{slm}, a balancing method \cite{balance}, an active constellation extension method \cite{ace}, a tone injection method \cite{tone}, an iterative filtering method \cite{iter} and a compounding method \cite{airy}. These methods are summarized in \cite{overview} \cite{litsyn}. In some of these methods, it is necessary to transmit some parameters as side information since receivers have to know the parameters to recover symbols. 

One of methods to reduce PAPR is a Partial Transmit Sequence (PTS) technique \cite{pts}-\cite{chicago}. PTS techniques are to multiply symbols by components to reduce PAPR. Therefore, it is necessary to transmit the vector as side information to the receiver. Further, with PTS techniques, OFDM signals are not distorted since we only modulate symbols. Then, their side lobes stay unchanged.

With PTS techniques, there is a significant task for reducing PAPR, how to reduce the calculation amount. In \cite{pts}, the suitable vector is chosen as one which achieves the lowest PAPR from all of the candidates. Then, the calculation amount exponentially gets larger as the length of a vector increases. To reduce such calculations, there are some methods. In \cite{neighborhood}, the neighborhood search algorithm has been proposed. With this method, we can obtain a local optimal solution. Another method is a phase random method \cite{phaserandom}. This method consists of generating random vectors whose phase is uniformly distributed in the set of the candidates. 

In this paper, we propose a method to search the vector which achieves low PAPR. The main point of our method is to obtain the vector from a set of random vectors generated from the Gaussian distribution. Therefore, our method is similar to a phase random method \cite{phaserandom}. Then, we derive the optimization problem to reduce PAPR and obtain a solution from the relaxed problem. We regard the solution as a covariance matrix and we can determine the Gaussian distribution. 

This paper is organized as follows: Section II shows the definitions of PAPR and Peak-to-Mean Envelope Power Ratio (PMEPR). In the literatures, these two notions sometimes are assumed to coincide, however, these two are different since PAPR and PMEPR are defined with RF signals and base-band signals, respectively. In this Section, we clarify the property of signals considered in this paper. Section III shows a partial transmit sequence technique and an optimization problem. It is not straightforward to solve this optimization problem since the feasible region of this optimization problem is discrete. Therefore, in Section IV, we show a semidefinite relaxation technique to solve it. With this technique, we can obtain approximate solutions. In Section V, we consider random vectors generated from the Gaussian distribution whose covariance matrix is the solution of the relaxed problem. Then, in Section VI, we show the relation between our randomization method and a phase random method, which is a conventional method. In Section VII, since our problem stated in Sections IV and V has the large number of constraints, we propose another optimization problem to reduce the upper bound of PAPR. In this problem, the number of constraints is less than one of that in Section IV and V. Finally, we compare the PAPR of our method with one of an existing phase random method. 


\section{OFDM System Model and PAPR}
In this section, we fix the model and the quantities used throughout this paper. A complex baseband OFDM signal is written as \cite{ofdmcdma}
\begin{equation}
  s(t) = \sum_{k=1}^{K} A_k \exp\left(2 \pi j \frac{k-1}{T}t\right), \hspace{2mm} 0 \leq t < T,
  \label{eq:ofdm}
\end{equation}
where $A_k$ is a transmitted symbol, $K$ is the number of symbols, $j$ is the unit imaginary number and $T$ is a duration of symbols. It is known that OFDM signals are generated by Inverse Fast Fourier Transformation (IFFT) \cite{ofdmcdma}. As seen in Eq. (\ref{eq:ofdm}), our OFDM signals have no cyclic prefixes. With the cyclic prefix technique, the PAPR of OFDM signals is preserved since the cyclic prefix does not introduce any new peaks \cite{sharif}. Therefore, we consider such OFDM signals written in Eq. (\ref{eq:ofdm}).

A Radio Frequency (RF) OFDM signal $\zeta(t)$ is written with Eq. (\ref{eq:ofdm}) as
\begin{equation}
  \begin{split}
    \zeta(t) &= \operatorname{Re}\{s(t)\exp(2 \pi j f_c t)\}\\
    &= \operatorname{Re}\left\{\sum_{k=1}^{K} A_k \exp\left(2 \pi j \left(\frac{k-1}{T} + f_c \right)t\right)\right\},
    \end{split}
\end{equation}
where $\operatorname{Re}\{z\}$ is the real part of $z$, and $f_c$ is a carrier frequency. With RF signals, PAPR is defined as \cite{compute} \cite{exist}
\begin{equation}
  \operatorname{PAPR} = \max_{0 \leq t < T}\frac{\left|\operatorname{Re}\left\{\displaystyle\sum_{k=1}^{K} A_k \exp\left(2 \pi j \left(\frac{k-1}{T} + f_c \right)t\right)\right\}\right|^2}{P_{\operatorname{av}}},
  \label{eq:PAPR}
\end{equation}
where $P_{\operatorname{av}}$ corresponds to the average power of signals, $P_{\operatorname{av}} = \sum_{k=1}^{K}\operatorname{E}\{|A_k|^2\}$, and $\operatorname{E}\{X\}$ is the average of $X$. Similarly, with baseband signals, PMEPR is defined as \cite{exist} \cite{compute}
\begin{equation}
  \operatorname{PMEPR} = \max_{0 \leq t < T}\frac{\left|\displaystyle\sum_{k=1}^{K} A_k \exp\left(2 \pi j \frac{k-1}{T} t\right) \right|^2}{P_{\operatorname{av}}}.
  \label{eq:PMEPR}
\end{equation}
In the literatures, PAPR and PMEPR have often been evaluated as probabilities, since PAPR and PMEPR depend on symbols $A_k$ that can be regarded as random variables \cite{ochiai} \cite{general}.

Obviously, PAPR does not always correspond to PMEPR. Further, from Eqs. (\ref{eq:PAPR}) and (\ref{eq:PMEPR}), PAPR does not exceed PMEPR. In \cite{sharif}, under some conditions described below, it has been proven that the following relations are established
\begin{equation}
  \left( 1 - \frac{\pi^2K^2}{2r^2}\right)\cdot \operatorname{PMEPR} \leq  \operatorname{PAPR} \leq  \operatorname{PMEPR},
  \label{eq:rel1}
\end{equation}
where $r$ is an integer such that $f_c = r/T$. The conditions that Eq. (\ref{eq:rel1}) holds are $K \ll r$ and $\exp(2 \pi j K/r) \approx 1$. In addition to these, another relation has been shown in \cite{litsyn}. Equation (\ref{eq:rel1}) implies that PMEPR approximately equals PAPR for sufficiently large $f_c$. It is often the case that PMEPR is evaluated instead of PAPR \cite{ochiai}. In what follows, we assume that the carrier frequency $f_c$ is sufficiently large, that is, we consider baseband OFDM signals instead of RF signals.  

\section{Partial Transmit Sequence Technique}
With OFDM systems, the Partial Transmit Sequence (PTS) technique has been proposed to reduce PAPR. In this section, we show the model and the details of PTS techniques. PTS techniques need a vector to reduce PAPR. A disadvantage of PTS techniques is that the large amount of calculation is necessary in some situations. The details of PTS techniques are described in \cite{litsyn} \cite{pts} \cite{pts2}.

Our symbols and how to derive an index sets for symbols are given as follows. We assume that the symbols $A_k$ are given and the number of the symbols is $K$. Further, the index set $\Lambda=\{1,2,\ldots,K\}$ corresponds to the set of unordered symbols $\{A_1,A_2,\ldots,A_{K}\}$. To apply a PTS technique, we divide the index set $\Lambda$ into $P$ disjoint subsets, $\Lambda_1,\ldots,\Lambda_{P}$, that is,
\begin{equation}
  \Lambda = \Lambda_1 \cup \cdots \cup \Lambda_{P}, \hspace{2mm}\Lambda_k \cap \Lambda_m = \emptyset\hspace{2mm}\mbox{if}\hspace{2mm}k \neq m
\end{equation}
for $k,m = 1,2,\ldots,P$. There are some discussions about how to divide the index set. We refer the reader to \cite{mseq} \cite{radix} \cite{random}. 

To express the instantaneous power, we define some quantities as below. For each subsets of symbols, we introduce a rotation vector $\mathbf{b} = (b_1,b_2,\ldots,b_{P})^\top$, where $\mathbf{x}^\top$ is the transpose of $\mathbf{x}$. The vector $\mathbf{b}$ is chosen as one satisfying $b_p = \exp(j \theta_p)$ for $p = 1,2,\ldots,P$, where $\theta_p \in [0,2\pi)$. This $\mathbf{b}$ plays various roles throughout this paper. For convenience, let us define the quantities
\begin{equation}
  A_k^{(p)} = \left\{ \begin{array}{cc}
                        A_k & k \in \Lambda_p\\
                        0 & \mbox{otherwise}
                      \end{array}
                    \right. 
\end{equation}
for $k = 1,2,\ldots,K$ and $p = 1,2,\ldots,P$. With $A_k^{(p)}$, a modified baseband OFDM signal $\hat{s}(t)$ is written as
\begin{equation}
  \hat{s}(t) = \sum_{p=1}^{P} \sum_{k=1}^{K} A_k^{(p)}b_p \exp\left(2 \pi j \frac{k-1}{T}t \right).
  \label{eq:pts}
\end{equation}
Note that the average power of modified signals is equivalent to one of the original OFDM signals since $|b_p|=1$. 
With a matrix and vectors, the above equation is rewritten as
\begin{equation}
  \hat{s}(t) = \mathbf{v}^\top_t A \mathbf{b},
\end{equation}
where
\begin{equation}
  \begin{split}
    \mathbf{v}_t &= \left(\begin{array}{cccc}
                          v_{1,t} & v_{2,t} & \cdots & v_{K,t}\\
                        \end{array}
                      \right)^\top,\\
                                            A &= \left( \begin{array}{cccc}
                                    A_1^{(1)} &  A_1^{(2)} & \cdots &  A_1^{(P)}\\
                                    A_2^{(1)} &  A_2^{(2)} & \cdots &  A_2^{(P)}\\
                                    \vdots & \vdots & \ddots & \vdots \\
                                    A_{K}^{(1)} &  A_{K}^{(2)} & \cdots &  A_{K}^{(P)}
                                  \end{array}
                                \right)\\
  \end{split}
\end{equation}
and
\begin{equation}
             v_{k,t} = \exp\left(2 \pi j \frac{k-1}{T}t \right).
\end{equation}
With these quantities, the instantaneous power $|\hat{s}(t)|^2$ is written as
\begin{equation}
  |\hat{s}(t)|^2 = \mathbf{b}^*A^*(\mathbf{v}_t^*)^\top\mathbf{v}^\top_t A \mathbf{b},
  \label{eq:inst}
\end{equation}
where $\mathbf{z}^*$ is the complex conjugate transpose of $\mathbf{z}$. We denote by $C_t$ the matrix $A^*(\mathbf{v}_t^*)^\top\mathbf{v}^\top_t A$. Note that $C_t$ is a positive semidefinite matrix since $C_t$ is a Gram matrix and each value of $b_p$ is chosen as one achieving the lowest PAPR.

At the receiver side, to recover symbols, it is necessary for the receiver to know the explicit values of $\mathbf{b}$. Note that Signal to Noise Ratio (SNR) is preserved if the receiver knows $\mathbf{b}$. To let the receiver know, the vector $\mathbf{b}$ has to be transmitted as side-information. To reduce information content, the value of $b_p$ is usually restricted to
\begin{equation}
  b_p \in \left\{1,\exp\left(2 \pi j \frac{1}{L}\right),\ldots,\exp\left(2 \pi j \frac{L-1}{L}\right)\right\},
\end{equation}
where $L$ is a positive integer. We denote by $\Omega_L$ the set $\left\{1,\exp\left(2 \pi j \frac{1}{L}\right),\ldots,\exp\left(2 \pi j \frac{L-1}{L}\right)\right\}$.  From the above definition, the vector $\mathbf{b} \in \Omega_L^P$, where $\Omega_L^P$ is the set of $P$-dimensioned vectors whose elements are in $\Omega_L$. Then, the information content of $\mathbf{b}$ is $(P-1) \log_2 L$ [bits] since we can set $b_1 = 1$ without loss of generality. It is obvious that the number of elements in the set $\Omega_L^{P-1}$ is $L^{P-1}$. Let $\mathbf{b}^\star$ be the vector which realizes the minimum PAPR. Then, it turns out that $\mathbf{b}^\star$ is the global solution of the optimization problem
\begin{equation}
  \begin{split}
    (Q_L) \hspace{3mm} & \hspace{3mm}\min \max_{0 \leq t < T} |\hat{s}(t)|^2\\
    \mbox{subject to}\hspace{3mm} & b_p \in \Omega_L \hspace{3mm} (p=1,2,\ldots,P).
  \end{split}
\end{equation}
Our aim is to find the vector $\mathbf{b}^\star$. To this end, there are two main obstacles to solve the problem $(Q_L)$. 

One obstacle is that the time $t$ is continuous. In \cite{sharif}, with baseband OFDM signals $s(t)$ defined in Eq. (\ref{eq:ofdm}), it has been shown that there is a following relation between continuous signals and sampled signals
\begin{equation}
  \max_{0 \leq t < T}|s(t)| < \sqrt{\frac{J^2}{J^2 - \pi^2/2}}\max_{0 \leq n < JK} \left|s\left(\frac{nT}{JK}\right)\right|,
  \label{eq:sample}
\end{equation}
where $J$ is an integer satisfying $J > \pi/\sqrt{2}$. Equation (\ref{eq:sample}) implies that PMEPR can be estimated precisely from signals sampled with a sufficiently large oversampling factor. For maxima of continuous signals and sampled signals, other relations have been shown in \cite{local} \cite{tellambura}. The integer $J$ is often called oversampling factor \cite{clip}. As an oversampling factor, $J \geq 4$ is often chosen. How to choose the oversampling factor $J$ has been discussed in \cite{sharif}.

With sampled signals, the problem $(Q_L)$ is rewritten as
\begin{equation}
  \begin{split}
    (\hat{Q}_L) \hspace{3mm} & \hspace{3mm} \min \lambda \\
    \mbox{subject to}\hspace{3mm} &  \mathbf{b}^* C_{nT/(JK)} \mathbf{b} \leq \lambda \hspace{2mm} (n = 0,1,\ldots,JK-1)\\
    & b_p \in \Omega_L \hspace{3mm} (p=1,2,\ldots,P)\\
    & \lambda \in \mathbb{R}.
  \end{split}
\end{equation}
Note that the variables in the problem $(\hat{Q}_L)$ are $\mathbf{b}$ and $\lambda$.

The other obstacle is that the feasible region $\Omega_L$ is discrete. In \cite{pts}, a brute-force search has been used to find the global solution $\mathbf{b}^\star$. With this method, we have to find the vector $\mathbf{b}^\star$ from $L^{P-1}$ candidates, and the calculation amount exponentially gets larger as $P$ increases. In \cite{neighborhood}, the neighborhood search algorithm has been proposed. With this method, we can obtain a local optimal solution. However, it is only known that its calculation amount is proportional to $_{P-1}C_r \cdot L^r$, where $r$ is an integer parameter expressing the distance of a neighborhood and $_aC_b$ is a binomial coefficient. Another existing method is a phase random method \cite{phaserandom}. This method consists of generating random vectors whose phase is uniformly distributed in the region $\Omega^P_L$, from which we obtain a solution. 

\section{Semidefinite Relaxation}
Since it is not straightforward to obtain the global solution, we propose an efficient method to obtain a solution which achieves low PAPR. Optimization problems, such as the problem $(\hat{Q}_L)$, appear in MIMO detection \cite{detect}. Thus, we can use these methods that have already been developed to our problem. One of such existing methods uses a semidefinite relaxation technique \cite{sdr}. In this section, we obtain a solution with such semidefinite relaxation techniques.

We apply semidefinite relaxation techniques to the problem $(\hat{Q}_L)$. Our main aim is to change the variable $\mathbf{b}$ to a positive semidefinite matrix $X$. The ways to solve the problem $(\hat{Q}_L)$ depend on $\Omega_L$. Therefore, we consider each problem for various cases of $L$.

\subsection{Optimization Problem for $L=2$}
First, we consider the problem $(\hat{Q}_L)$ for $L=2$, $(\hat{Q}_2)$. Then, $\Omega_2 = \{-1,1\}$. Note that $b^2=1$ for $b \in \Omega_2$. If we define the matrix $X = \mathbf{b}\mathbf{b}^\top$, then $X$ is a positive semidefinite matrix whose rank is 1 and the problem $(\hat{Q}_2)$ is rewritten as
\begin{equation}
  \begin{split}
    (\hat{Q}_2) \hspace{3mm} & \hspace{3mm} \min \lambda \\
    \mbox{subject to}\hspace{3mm} &  \operatorname{Tr}(C_{nT/(JK)}X) \leq \lambda \hspace{2mm} (n = 0,1,\ldots,JK-1)\\
    & X_{p,p} = 1 \hspace{3mm} (p=1,2,\ldots,P)\\
    & \operatorname{rank}(X) = 1\\
    & X \succcurlyeq 0\\
    & X \in \mathbb{S}_P, \qquad \lambda \in \mathbb{R},
  \end{split}
\end{equation}
where $\operatorname{Tr}(X)$ is the trace of $X$, $\operatorname{rank}(X)$ is the rank of $X$, $X \succcurlyeq 0$ indicates that $X$ is a positive semidefinite matrix and $\mathbb{S}_P$ is the set of symmetric matrices of dimension $P$. Due to the constraint $\operatorname{rank}(X) = 1$, the problem $(\hat{Q}_2)$ is not convex. Note that the set of positive semidefinite matrices is convex \cite{boyd}. By dropping the rank constraint, we obtain the relaxed optimization problem  
\begin{equation}
  \begin{split}
    (\hat{Q}'_2) \hspace{3mm} & \hspace{3mm} \min \lambda \\
    \mbox{subject to}\hspace{3mm} &  \operatorname{Tr}(C_{nT/(JK)}X) \leq \lambda \hspace{2mm} (n = 0,1,\ldots,JK-1)\\
    & X_{p,p} = 1 \hspace{3mm} (p=1,2,\ldots,P)\\
    & X \succcurlyeq 0\\
    & X \in \mathbb{S}_P, \qquad \lambda \in \mathbb{R}.
  \end{split}
\end{equation}
The above problem $(\hat{Q}'_2)$ can be immediately solved since the problem $(\hat{Q}'_2)$ is convex. Let $X^\star_2$ be the global solution of the problem $(\hat{Q}'_2)$.  If the rank of $X^\star_2$ is 1, then we obtain the global solution of the problem $(\hat{Q}_2)$, denoted by $\mathbf{b}^\star_2$. However, the rank of $X^\star_2$ is not always 1. To deal with this, we obtain an approximate solution from $X^\star_2$. In general, the solution $X^\star_2$ is decomposed as 
\begin{equation}
  X^\star_2 = \sum_{i=1}^{r_2} \lambda_i \mathbf{q}_i \mathbf{q}_i^*, 
\end{equation}
where $r_2 = \operatorname{rank}(X^\star_2)$, $\lambda_i$ is the eigenvalue of $X^\star_2$, $\lambda_1 \geq \lambda_2 \geq \ldots \geq \lambda_{r_2}$ and $\mathbf{q}_i$ is the respective eigenvector. Then, in a least two norm sense, the approximate solution whose rank 1 is obtained as $\hat{X}^\star_2 = \lambda_1 \mathbf{q}_1 \mathbf{q}_1^*$. From this approximate solution, we systematically obtain the solution of the original problem $(\hat{Q}_2)$ as $\sqrt{\lambda_1}\mathbf{q}_1$. However, this solution is not always in the feasible region of the problem $(\hat{Q}_2)$. To have an approximate solution in the feasible region, for the problem $(\hat{Q}_2)$, we need to project the solution onto the feasible region. We arrive at the $p$-th element of the approximate solution of the problem $(\hat{Q}_2)$ as
\begin{equation}
  \hat{b}_{2,p} = \operatorname{sgn}(q_{1,p}),
\end{equation}
where $q_{1,p}$ is the $p$-th element of $\mathbf{q}_1$ and
\begin{equation}
  \operatorname{sgn}(x) = \left\{ \begin{array}{c c}
                                    1 & x \geq 0\\
                                    -1 & x < 0
                                  \end{array}
                                  \right. .
\end{equation}

\subsection{Optimization Problem for $L=4$}
Similarly, for $L=4$, we obtain the approximate solution of the problem $(\hat{Q}_4)$. For $L=4$, the set $\Omega_4$ is written as $\Omega_4 = \{1, \exp(j \pi/2),-1,\exp(j 3\pi/4)\}$. To obtain the relaxed problem, we rewrite the problem $(\hat{Q}_4)$ as follows. First, let us define the set
\begin{equation}
  \hat{\Omega}_4 = \{+1+j,+1-j,-1+j,-1-j\},
\end{equation}
which can be expressed as
\begin{equation}
  \hat{\Omega}_4 = \{\sqrt{2}\exp(j \pi/4) \cdot a \mid a \in \Omega_4\}.
\end{equation}
Note that $\operatorname{Re}\{b\}^2=\operatorname{Im}\{b\}^2=1$ for $b \in \hat{\Omega}_4$. Second, since the set $\hat{\Omega}_4$ consists of complex elements, we rewrite the set $\hat{\Omega}_4$ in terms of real parts and imaginary parts. We introduce the following transformations for $\mathbf{z} \in \mathbb{C}^n$ and $Z \in \mathbb{H}_n$, where $\mathbb{H}_n$ is the set of Hermitian matrices of dimension $n$ \cite{maxcut},
\begin{equation*}
  \mathcal{T}(\mathbf{z}) = \left( \begin{array}{c}
                                     \operatorname{Re}\{\mathbf{z}\}\\
                                     \operatorname{Im}\{\mathbf{z}\}
                                   \end{array} \right),\hspace{3mm}\mbox{and}\hspace{3mm}
                                 \mathcal{T}(Z) = \left( \begin{array}{cc}
                                     \operatorname{Re}\{Z\} & -\operatorname{Im}\{Z\}\\
                                     \operatorname{Im}\{Z\} & \operatorname{Re}\{Z\}
                                   \end{array} \right).
\end{equation*}
Note that $\mathcal{T}(X) \in \mathbb{S}_{2n}$ if $X \in \mathbb{H}_n$ \cite{goemans}. Finally, with the above operations, we arrive at the relaxed problem for $L=4$.
\begin{equation}
  \begin{split}
    (\hat{Q}'_4) \hspace{3mm} & \hspace{3mm} \min \lambda \\
    \mbox{subject to}\hspace{3mm} &  \operatorname{Tr}(\hat{C}_{nT/(JK)}X) \leq \lambda \hspace{2mm} (n = 0,1,\ldots,JK-1)\\
    & X_{p,p} = 1 \hspace{3mm} (p=1,2,\ldots,2P)\\
    & X \succcurlyeq 0\\
    & X \in \mathbb{S}_{2P}, \qquad \lambda \in \mathbb{R},
  \end{split}
\end{equation}
where $\hat{C}_t = \mathcal{T}(C_t)$. Note that the problem $(\hat{Q}'_4)$ is equivalent to $(\hat{Q}_4)$ if we impose the rank constraints to the problem $(\hat{Q}'_4)$ and the problem $(\hat{Q}'_4)$ is convex. Let $X^\star_4$ be the global solution of the problem $(\hat{Q}'_4)$. From $X^\star_4$, we can obtain an approximate solution as follows. Similar to the problem for $L=2$, $X^\star_4$ is decomposed as
\begin{equation}
  X^\star_4 = \sum_{i=1}^{r_4} \lambda_i \mathbf{q}_i \mathbf{q}_i^*, 
\end{equation}
where $r_4 = \operatorname{rank}(X^\star_4)$, $\lambda_i$ is the eigenvalue of $X^\star_4$, $\lambda_1 \geq \lambda_2 \geq \ldots \geq \lambda_{r_4}$ and $\mathbf{q}_i$ is the respective eigenvector. From the vector $\mathbf{q}_1$, we can obtain the approximate solution $\hat{\mathbf{b}}_4 \in \mathbb{C}^P$ written as
\begin{equation}
  \hat{b}_{4,p} = \frac{1}{\sqrt{2}}\exp(-j \pi/4)\left(\operatorname{sgn}(q_p) + j \cdot \operatorname{sgn}(q_{p+P})\right),                          
\end{equation}
where $\hat{b}_{4,p}$ and $q_p$ are the $p$-th elements of $\hat{\mathbf{b}}_4$ and $\mathbf{q}_1$, respectively.

\subsection{Optimization Problem for General $L$}
For general $L$, we consider the relaxed problem
\begin{equation}
  \begin{split}
    (\hat{Q}'_L) \hspace{3mm} & \hspace{3mm} \min \lambda \\
    \mbox{subject to}\hspace{3mm} &  \operatorname{Tr}(C_{nT/(JK)}X) \leq \lambda \hspace{2mm} (n = 0,1,\ldots,JK-1)\\
    & X_{p,p} = 1 \hspace{3mm} (p=1,2,\ldots,P)\\
    & X \succcurlyeq 0\\
    & X \in \mathbb{H}_P, \qquad \lambda \in \mathbb{R}.
  \end{split}
\end{equation}
Note that the set of Hermitian semidefinite positive matrices is convex \cite{sdr} and the above problem $(\hat{Q}'_L)$ is convex. The problem $(\hat{Q}'_L)$ is not equivalent to the problem $(\hat{Q}_L)$ for $L \neq 2,4$ if the rank constraint is imposed. Similar to the problem for $L=2$ and $L=4$, the approximate solution can be obtained as follows. Let $X^\star_L$ be the global solution of the problem $(\hat{Q}'_L)$. Then, $X^\star_L$ is decomposed as
\begin{equation}
  X^\star_L = \sum_{i=1}^{r_L} \lambda_i \mathbf{q}_i \mathbf{q}_i^*, 
\end{equation}
where $r_L = \operatorname{rank}(X^\star_L)$, $\lambda_i$ is the eigenvalue of $X^\star_L$, $\lambda_1 \geq \lambda_2 \geq \ldots \geq \lambda_{r_L}$ and $\mathbf{q}_i$ is the respective eigenvector. From the vector $\mathbf{q}_1$, we can obtain the approximate solution $\hat{\mathbf{b}}_L \in \mathbb{C}^P$ as
\begin{equation}
  \hat{\mathbf{b}}_L = \argmin_{\mathbf{b} \in \Omega_L^P} \|\mathbf{q} - \mathbf{b}\|,
\end{equation}
where $\mathbf{q} = \sqrt{\lambda_1}\mathbf{q}_1$ and $\|\mathbf{z}\|$ is the Euclidean norm of $\mathbf{z}$.

\section{Randomization Method}
In Section IV, we have discussed the relaxed problems and how to obtain the approximate solutions. However, clearly, approximate solutions are not suitable if the global solutions of the relaxed problems have some large eigenvalues, that is, the ranks of solutions are not regarded as unity. 

In this section, we introduce a randomization method. This method is used to analyze how far the optimal value of relaxed problems is from one of original problems \cite{randap}. With this method, we obtain solutions as random values which are generated from Gaussian distribution. Similar to discussions in Section IV, we consider each problem for various cases of $L$.

\subsection{Randomization for $L=2$}
First, we consider the problem for $L=2$. Let $\boldsymbol{\xi}$ be a random vector generated from the Gaussian distribution $\mathcal{N}(\mathbf{0},X)$ with zero mean and a covariance matrix $X$. The definition and properties of a Gaussian distribution have been shown in \cite{aspects}. 

To find an approximate solution, we rewrite the problem $(\hat{Q}'_2)$ as follows. With
\begin{equation}
  \operatorname{E}\{\boldsymbol{\xi}^\top C_t\boldsymbol{\xi}\} = \operatorname{Tr}(C_tX),
\end{equation}
the problem $(\hat{Q}'_2)$ can be written as
\begin{equation}
  \begin{split}
    (\hat{Q}'_2) \hspace{3mm} & \hspace{3mm} \min \lambda \\
    \mbox{subject to}\hspace{3mm} &  \operatorname{E}\{\boldsymbol{\xi}^\top C_{nT/(JK)}\boldsymbol{\xi}\} \leq \lambda \hspace{2mm} (n = 0,1,\ldots,JK-1)\\
    & X_{p,p} = 1 \hspace{3mm} (p=1,2,\ldots,P)\\
    & X \succcurlyeq 0\\
    & \boldsymbol{\xi} \sim \mathcal{N}(0,X)\\
    & X \in \mathbb{S}_P, \qquad \lambda \in \mathbb{R}.
  \end{split}
\end{equation}
Note that the variables of the above problem are $X$ and $\lambda$. Then, it is clear that the optimal matrix $X^\star_2$ defined in Section IV is the optimal matrix of the above problem in a sense of a covariance matrix. This result suggests that a suitable solution can be obtained from a set of random vectors generated from the Gaussian distribution $\mathcal{N}(\mathbf{0},X^\star_2)$ \cite{sdprand}. We can then obtain the approximate solution as follows. 
\begin{enumerate}
\item Solve the problem $(\hat{Q}'_2)$ and obtain the covariance matrix $X^\star_2$.
\item Generate random vectors $\{\boldsymbol{\xi}\}$ from the Gaussian distribution $\mathcal{N}(\mathbf{0},X^\star_2)$ and project them onto the feasible region of the original problem $(\hat{Q}_L)$, that is, for $L=2$, obtain the projected solutions
\begin{equation}
  \hat{b}_p = \operatorname{sgn}\left(\xi_p\right) \hspace{2mm}(p=1,2,\ldots,P),
\end{equation}
where $\xi_p$ is the $p$-th element of $\boldsymbol{\xi}$.
\item Choose the solution which achieves the minimum PAPR among all the random vectors and regard it as an approximate solution.
\end{enumerate}

\subsection{Randomization for $L=4$}
Similar to the case for $L=2$, we can obtain the covariance matrix $X^\star_4$ for $L=4$ and obtain random vectors $\{\boldsymbol{\xi}\}$ generated from $\mathcal{N}(\mathbf{0},X^\star_4)$. Since the dimension of the vectors $\{\boldsymbol{\xi}\}$ is $2P$, the way to project is written as
\begin{equation}
\hat{b}_p  = \frac{1}{\sqrt{2}}\exp(-j \pi/4)\left(\operatorname{sgn}(\xi_p) + j \cdot \operatorname{sgn}(\xi_{p+P})\right)  
\end{equation}
for $p=1,2,\ldots,P$. From these random vectors, we choose an approximate solution which achieves the minimum PAPR among them.

\subsection{Randomization with General $L$}
For general $L$, we can obtain the complex covariance matrix $X^\star_L$ as a solution of the problem $(\hat{Q}'_L)$. Similar to the methods for $L=2$ and $L=4$, our goal is to choose the solution from random vectors. Our main part of our method is to obtain an approximate solution from random vectors generated from the complex Gaussian distribution $\mathcal{CN}(\mathbf{0},X^\star_L)$. The definition and the detail of a complex Gaussian distribution have been shown in \cite{graphical}. There are some methods to obtain an approximate solution from random vectors \cite{proj1}-\cite{proj3}, and our method is a special case of an algorithm in \cite{proj3}. From the complex Gaussian distribution $\mathcal{CN}(\mathbf{0},X^\star_L)$, we can obtain the random vectors $\{\boldsymbol{\xi}\}$. Then, we have to transform the random vectors $\{\boldsymbol{\xi}\}$ into feasible ones as solutions of the problem $(\hat{Q}_L)$. Our transformation method is written as follows. Let $f_L$ be
\begin{equation}
  f_L(z) = \left\{ \begin{array}{c l}
                     1 & \operatorname{Arg} z \in [0,\frac{1}{L}2\pi)\\
                     \omega_L & \operatorname{Arg} z \in [\frac{1}{L}2\pi,\frac{2}{L}2\pi)\\
                     \vdots & \\
                     \omega^l_L & \operatorname{Arg} z \in [\frac{l}{L}2\pi,\frac{l+1}{L}2\pi)\\
                     \vdots & \\
                     \omega^{L-1}_L & \operatorname{Arg} z \in [\frac{L-1}{L}2\pi,2\pi)
                   \end{array}
                   \right. ,
\end{equation}
where $z \in \mathbb{C}$, $\omega_L = \exp(2 \pi j / L)$ and $\operatorname{Arg} z$ is the angle of $z$. With the function $f_L$, the random vector $\boldsymbol{\xi}$ generated from the complex Gaussian distribution $\mathcal{CN}(\mathbf{0},X^\star_L)$ is transformed to
\begin{equation}
  \hat{b}_p = f_L(\xi_p) \hspace{2mm}(p=1,2,\ldots,P).
\end{equation}
It is clear that $\hat{b}_p \in \Omega_L$. Therefore, $\hat{\mathbf{b}}$ is a feasible solution of the problem $(\hat{Q}_L)$. With the above method, we can obtain the feasible solutions from random vectors generated from the complex Gaussian distribution $\mathcal{CN}(\mathbf{0},X^\star_L)$. Then, we choose the approximate solution from them which achieves the minimum PAPR among the set of the random vectors. Our method is summarized as follows,
\begin{algorithm}[h]
Obtain the relaxed problem with semidefinite relaxation techniques.\\
Obtain the positive semidefinite matrix $X^\star$ as the optimal solution of the relaxed problem.\\
Determine the Gaussian distribution $\mathcal{N}(\mathbf{0},X^\star)$ (with $L \neq 2,4$, $\mathcal{CN}(\mathbf{0},X^\star)$ is determined). Then, generate $N$ samples from the Gaussian distribution as the candidates of the solution.\\
Project the samples onto the feasible region, and obtain the projected samples.\\
Choose the solution $\mathbf{b}^\star$ from the projected samples which achieves the minimum PAPR. Then, output $\mathbf{b}^\star$ as the solution.
\caption{Randomization Method with Semidefinite Relaxation}
\label{algo:ours}
\end{algorithm}

\section{Relation between Our Method and Phase Random Method}
In Section V, we have shown our randomization method. Similar to our method, a phase random method has been proposed \cite{phaserandom}. This method uses random vectors whose phase is uniformly distributed in $\Omega_L$. In this Section, we discuss the relation between our method and a phase random method. 

First, we explain a phase random method. We define the probability mass function as
\begin{equation}
  \operatorname{Pr}\left\{z = \omega^l_L\right\} = \frac{1}{L}\hspace{2mm}(l=0,1,\ldots,L-1),
\end{equation}
where $\omega_L = \exp(2 \pi j /L)$, defined in Section V. Then, $\omega_L \in \Omega_L$ and phases are uniformly distributed in $\Omega_L$.

Further, let us discuss the complex Gaussian distribution. From \cite{graphical}, if $\mathbf{z} \in \mathbb{C}^n$ follows $\mathcal{CN}(\boldsymbol{\mu},\Sigma)$, then the probability density function of $\mathcal{T}(\mathbf{z}) \in \mathbb{R}^{2n}$ is the Gaussian distribution $\mathcal{N}\left(\mathcal{T}(\boldsymbol{\mu}),\frac{1}{2}\mathcal{T}(\Sigma)\right)$. Therefore, we can consider a real-value Gaussian distribution instead of a complex Gaussian distribution.

Let us consider the complex Gaussian distribution $\mathcal{CN}(\mathbf{0},I_P)$, where $I_P$ is the identity matrix whose size is $P$. It is clear that the matrix $\mathcal{T}(I_P)$ is the identity matrix whose size is $2P$. From the above discussion, and the covariance matrix is identity matrix, each variable of $\mathbf{z}$ generated from $\mathcal{CN}(\mathbf{0},I_P)$ is uncorrelated. It is known in \cite{prob} that uncorrelatedness is equivalent to independence for normal variables. Therefore, it is sufficient to consider a vector $\mathbf{z}$ whose element is generated from the complex Gaussian distribution $\mathcal{CN}(0,1)$. The variable $z$ which is the element of $\mathbf{z}$ can be decomposed as
\begin{equation}
z = x + jy,
\end{equation}
where $x$ and $y$ are real numbers following the independent Gaussian distribution $\mathcal{N}(0,1/2)$, respectively.

Let us define $r \geq 0$ and $\theta \in [0,2\pi)$ so that
\begin{equation}
x + jy = r \exp(j\theta)
\end{equation}
Then, since $x$ and $y$ are normal variables following $\mathcal{N}(0,1/2)$, the probability density of $\theta \in [0,2\pi]$ is \cite{prob}
\begin{equation}
  p(\theta) = \frac{1}{2\pi},
\end{equation}
from which, the phase of a variable $z$ generated from $\mathcal{CN}(0,1)$ is uniformly distributed.

From the above discussions and the definition of the function $f_L(z)$, the probability mass function of $f_L(z)$ is written as
\begin{equation}
  \operatorname{Pr}\left\{f_L(z) = \omega^l_L\right\} = \frac{1}{L}.
\end{equation}
This result implies that a phase random method is equivalent to our method whose covariance matrix is the identity matrix with the function $f_L(z)$.

\section{Reducing Upper Bound of PAPR}
We have discussed how to obtain a covariance matrix to determine a Gaussian distribution. In Section V, we have obtained the optimization problem $(\hat{Q}'_L)$. This problem contains the oversampling parameter $J$. As seen in Section III, measured PAPR calculated from sampled signals converges to the true value of PAPR as $J \rightarrow \infty$. Therefore, a sufficiently large $J$ is necessary to evaluate PAPR tightly. Then, however, the number of constraints in the optimization problem $(\hat{Q}'_L)$ gets larger as $J$ increases. In such a situation, the optimization problem $(\hat{Q}'_L)$ gets complicated.

To overcome this obstacle, instead of PAPR, we consider an optimization problem to reduce the upper bound of PAPR which does not depend on time $t$. From this problem, we obtain a covariance matrix as the solution.

In this Section, we consider a general $L$. Then, specifying $L=2,4$, we can verify the same results to ones obtained in this Section with the techniques discussed in Section IV: with $L=2$, the set of matrices is the symmetric matrices $\mathbb{S}_P$, and with $L=4$, we replace a positive semidefinite matrix $X$ with $\mathcal{T}(X)$. 

The upper bound of the signal envelope has been shown with Eq. (\ref{eq:ofdm}) as \cite{tellambura_bound}
\begin{equation}
  |s(t)|^2 \leq \sum_{k=1}^K|A_k|^2 + 2\sum_{i=1}^{K-1}|\rho(i)|,
  \label{eq:bound_PAPR}
\end{equation}
where
\begin{equation}
  \rho(i) = \sum_{k=1}^{K-i}A_k\overline{A}_{k+i}
\end{equation}
and $\overline{z}$ is the complex conjugate of $z$.
We define $\rho(K)=0$. The right hand side of Eq. (\ref{eq:bound_PAPR}) is independent of the time $t$. Let us define $\boldsymbol{\rho'}=(\rho(1),\rho(2),\ldots,\rho(K-1))^\top$. Note that the first term in right side of Eq. (\ref{eq:bound_PAPR}), $\sum_{k=1}^K|A_k|^2$ corresponds to $\rho(0)$ and this term is not varied with PTS techniques since each element of a vector $b_n$ satisfies $|b_n|=1$.

From the above discussion, without taking into account convexity, it is expected to decrease PAPR when we reduce $\|\boldsymbol{\rho'}\|_{l_1}$, where $\|\mathbf{z}\|$ is the $l_1$-norm of $\mathbf{z}$. However, it is not the case since each $|\rho(i)|$ is not convex if we regard $A_k$ as variables. Therefore, we use $l_2$-norm of $\boldsymbol{\rho'}$, $\|\boldsymbol{\rho'}\|_{l_2}$. From the Cauchy-Schwarz inequality, it follows that
\begin{equation}
  \|\boldsymbol{\rho'}\|_{l_1} \leq \sqrt{K-1}\|\boldsymbol{\rho'}\|_{l_2}.
\end{equation}
Therefore, $\|\boldsymbol{\rho'}\|_{l_1}$ is expected to be reduced when $\|\boldsymbol{\rho'}\|_{l_2}$ is reduced.

Let us consider the vector $\hat{\boldsymbol{\rho}}=(\rho(0),\sqrt{2}\rho(1),\ldots,\sqrt{2}\rho(K-1))^\top$. It is clear that minimizing $\|\hat{\boldsymbol{\rho}}\|_{l_2}$ is equivalent to minimizing $\|\sqrt{2}\boldsymbol{\rho'}\|_{l_2}$ since $\rho(0)$ is constant. Then, $\|\hat{\boldsymbol{\rho}}\|^2_{l_2}$ is written as
\begin{equation}
  \begin{split}
   \| \hat{\boldsymbol{\rho}}\|^2_{l_2} =& 2\sum_{k=1}^{K-1}|\rho(k)|^2 + |\rho(0)|^2\\
    =& \sum_{k=0}^{K-1}|\rho(k)|^2 + \sum_{k=0}^{K-1}|\rho(K-k)|^2 \\
    =&\frac{1}{2}\left\{\sum_{k=0}^{K-1}|\rho(k) + \overline{\rho(K-k))}|^2\right.\\
    &\left. + \sum_{k=0}^{K-1}|\rho(k) - \overline{\rho(K-k)}|^2\right\}.
  \end{split}
    \label{eq:l2_expression}
\end{equation}
From the above equations, $\|\hat{\boldsymbol{\rho}}\|^2_{l_2}$ is divided into a periodic correlation term and an odd periodic correlation term. With Eq. (\ref{eq:pts}), these terms are written as
\begin{equation}
  \begin{split}
    \rho(k) + \overline{\rho(K-k)} &= \mathbf{b}^*A^* B^{(k)}_{1,1}A\mathbf{b}\\
    \rho(k) - \overline{\rho(K-k)} &= \mathbf{b}^*A^* B^{(k)}_{-1,1}A\mathbf{b},\\
  \end{split}
\end{equation}
where the matrices $B^{(k)}_{1,1}$ and $B^{(k)}_{-1,1}$ are written as
\begin{equation}
    B^{(k)}_{1,1} = \left( \begin{array}{cc}
                       O & I_{k}\\
                       I_{K-k} & O
                      \end{array} \right),\hspace{3mm}
   B^{(k)}_{-1,1} = \left( \begin{array}{cc}
                       O & -I_{k}\\
                       I_{K-k} & O
                             \end{array} \right).
\end{equation}
Since these matrices are regular matrices, they can be transformed to diagonal matrices. With this general discussion, these matrices are decomposed as \cite{mypaper}
\begin{equation}
  \begin{split}
    B^{(k)}_{1,1} &= V^*D^{(k)}V\\
   B^{(k)}_{-1,1} &= \hat{V}^*\hat{D}^{(k)}\hat{V},
  \end{split}
\end{equation}
where $V$ and $\hat{V}$ are unitary matrices whose $(m,n)$-th elements are
\begin{equation}
\begin{split}
V_{m,n} &= \frac{1}{\sqrt{K}}\exp\left(-2 \pi j \frac{mn}{K}\right),\\
 \hat{V}_{m,n} &= \frac{1}{\sqrt{K}}\exp\left(-2 \pi j n\left(\frac{m}{K} + \frac{1}{2K}\right)\right),
\end{split}
\end{equation}
and $D^{(k)}$ and $\hat{D}^{(k)}$ are diagonal matrices whose $n$-th diagonal elements are
\begin{equation}
\begin{split}
D_n^{(k)} &=  \exp\left(-2 \pi j k\frac{n}{K}\right),\\
  \hat{D}_n^{(k)} &=  \exp\left(-2 \pi j k\left(\frac{n}{K} + \frac{1}{2K}\right)\right).
\end{split}
\end{equation}
With these expressions, Eq. (\ref{eq:l2_expression}) is written as
\begin{equation}
  \|\hat{\boldsymbol{\rho}}\|_{l_2}^2 = \frac{K}{2}\left\{\sum_{k=1}^{K}|\alpha_k|^4 + \sum_{k=1}^{K}|\beta_k|^4\right\},
\end{equation}
where $\alpha_k$ and $\beta_k$ are the $k$-th element of $\boldsymbol{\alpha}$ and $\boldsymbol{\beta}$ written as $\boldsymbol{\alpha}=VA\mathbf{b}$ and $\boldsymbol{\beta} = \hat{V}A\mathbf{b}$, respectively.
With the variable $\mathbf{b}$, the above equation is written as
\begin{equation}
  \|\hat{\boldsymbol{\rho}}\|_{l_2}^2 = \frac{K}{2}\sum_{k=1}^K \left\{\left(\mathbf{b}^*A^*V^*G_kVA\mathbf{b}\right)^2 + \left(\mathbf{b}^*
      A^*\hat{V}^*G_k\hat{V}A\mathbf{b}\right)^2\right\},
  \label{eq:l2_expression2}
\end{equation}
where $G_k$ is a matrix whose $(k,k)$-th element is unity and the other elements are zero. Note that $G_k = G_k^*G_k$. Then, the matrices $A^*V^*G_kVA$ and $A^*\hat{V}^*G_k\hat{V}A$ are positive semidefinite matrices since they are the Gram matrices. Further, Eq. (\ref{eq:l2_expression2}) is convex with respect to the variable $\mathbf{b}$. This is proven in Appendix A. From the above discussions, it follows that the squared $l_2$ norm of $\hat{\boldsymbol{\rho}}$ is a convex function with respect to the variable $\mathbf{b}$. Combining these discussions above, we obtain the optimization problem,
\begin{equation}
  \begin{split}
    (Q_{l_2}) \hspace{3mm} & \hspace{3mm}\min F(\mathbf{b})\\
    \mbox{subject to}\hspace{3mm} & b_p \in \Omega_L \hspace{3mm} (p=1,2,\ldots,P),
  \end{split}
\end{equation}
where
\begin{equation}
  F(\mathbf{b})=\sum_{k=1}^K \left\{\left(\mathbf{b}^*A^*V^*G_kVA\mathbf{b}\right)^2 + \left(\mathbf{b}^*
      A^*\hat{V}^*G_k\hat{V}A\mathbf{b}\right)^2\right\}.
\label{eq:upper_obj}
\end{equation}
To deal the discrete set $\Omega_L$, we obtain the relaxed problem with semidefinite relaxation techniques. This convex problem is written as
\begin{equation}
  \begin{split}
    (Q'_{l_2}) \hspace{3mm} & \hspace{3mm}\min \hat{F}(X)\\
    \mbox{subject to}\hspace{3mm} & X_{p,p} = 1 \hspace{3mm} (p=1,2,\ldots,P)\\
    & X \succcurlyeq 0\\
    & X \in \mathbb{H}_P,
  \end{split}
\end{equation}
where
\begin{equation}
  \hat{F}(\mathbf{X})=\sum_{k=1}^K \left\{\left(A^*V^*G_kVAX\right)^2 + \left(
      A^*\hat{V}^*G_k\hat{V}AX\right)^2\right\}.
\label{eq:upper_relax_obj}
\end{equation}

From the above discussions, how to obtain the optimal solution as a positive semidefinite matrix is shown. Then, we discuss the relation between our randomization method and the relaxed problem $(Q'_{l_2})$. Let $X^\star$ and $\{\boldsymbol{\xi}\}$ be the global solution of the problem $(Q'_{l_2})$ and the random vectors generated from the Gaussian distribution $\mathcal{CN}(\mathbf{0},X^\star)$, respectively. They satisfy $\operatorname{E}\{\boldsymbol{\xi}\boldsymbol{\xi}^*\} = X^\star$. Then, the relations are shown
\begin{equation}
  \begin{split}
   & \sum_{k=1}^K \left\{\operatorname{Tr}\left(A^*V^*G_kVAX^\star\right)^2 + \operatorname{Tr}\left(
       A^*\hat{V}^*G_k\hat{V}AX^\star\right)^2\right\}\\
   \leq & \sum_{k=1}^K \operatorname{E}\left\{\left(\boldsymbol{\xi}^*A^*V^*G_kVA\boldsymbol{\xi}\right)^2 + \left(\boldsymbol{\xi}^*
       A^*\hat{V}^*G_k\hat{V}A\boldsymbol{\xi}\right)^2\right\}\\
   \leq & 3\sum_{k=1}^K \left\{\operatorname{Tr}\left(A^*V^*G_kVAX^\star\right)^2 + \operatorname{Tr}\left(
       A^*\hat{V}^*G_k\hat{V}AX^\star\right)^2\right\}.
  \end{split}
\label{eq:upper_relation}
\end{equation}
The above relations are proven in Appendix B.
Our main aim is to find $X^\star_{l_2}$ minimizing $\operatorname{E}\left\{F(\boldsymbol{\xi})\right\}$ under the constraints, where $\boldsymbol{\xi} \sim \mathcal{CN}(\mathbf{0},X^\star_{l_2})$. Two inequalities are involved in Eq. (\ref{eq:upper_relation}). The first inequality in Eq (\ref{eq:upper_relation}) implies that the global solution of the relaxed problem $X^\star$ does not always correspond to $X^\star_{l_2}$. However, the last inequality in Eq (\ref{eq:upper_relation}) implies that $X^\star$ will be an appropriate solution for our randomization method since $X^\star$ will makes $\operatorname{E}\left\{F(\boldsymbol{\xi})\right\}$ small where $\boldsymbol{\xi} \sim \mathcal{CN}(\mathbf{0},X^\star)$. From the above discussions, the global solution of the relaxed problem $X^\star$ is not the optimal covariance matrix with our randomization method minimizing upper bound of PAPR. However, $X^\star$ will achieve low PAPR with our randomization method.

\section{Numerical Results}
We numerically solve the problems $(\hat{Q}'_L)$ and $(Q'_{l_2})$ with CVX \cite{cvx} and obtain approximate solutions with the two kinds of methods, a $l_2$ approximation method discussed in Section IV and a random method discussed in Sections V and VII, respectively. As the parameters, the number of carriers $K=256$ and the oversampling parameter $J=16$ are chosen. We obtain PAPR curves with three kinds of parameters, $(P,L)=(16,2), (8,4)$ and $(8,8)$. The oversampling parameter $J$ is also used in calculating PAPR (see Eq. (\ref{eq:sample})). As the modulation scheme, each symbol is independently chosen from 16QAM symbols. The index sets $\Lambda_n$ are fixed and chosen as adjacent sets, that is
\begin{equation}
  \Lambda_n = \left\{\frac{K}{P}(n-1)+1,\ldots,\frac{K}{P}(n-1)+P\right\}
\end{equation}
for $n = 1,2,\ldots,P$. With our randomization methods discussed in Sections V and VII, and the phase random method, we generate 10 and 70 samples as solution candidates and choose the optimal solution from such candidates (see Algorithm \ref{algo:ours}). For the brute force method and the other methods, we draw the PAPR curves from 200 results and 2000 results, respectively. Note that the PAPR curve with the brute force method is optimal. 

\begin{figure}[htbp]
\centering  
\includegraphics[width=2.9in]{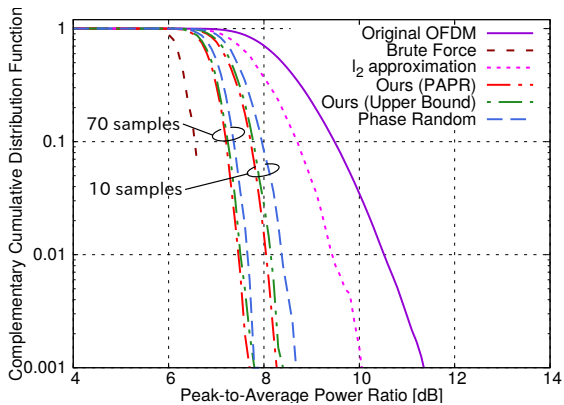}
\caption{PAPR with the parameters $(P,L)=(16,2)$}
\label{fig:P16L2}
\end{figure}

\begin{figure}[htbp]
\centering
\includegraphics[width=2.9in]{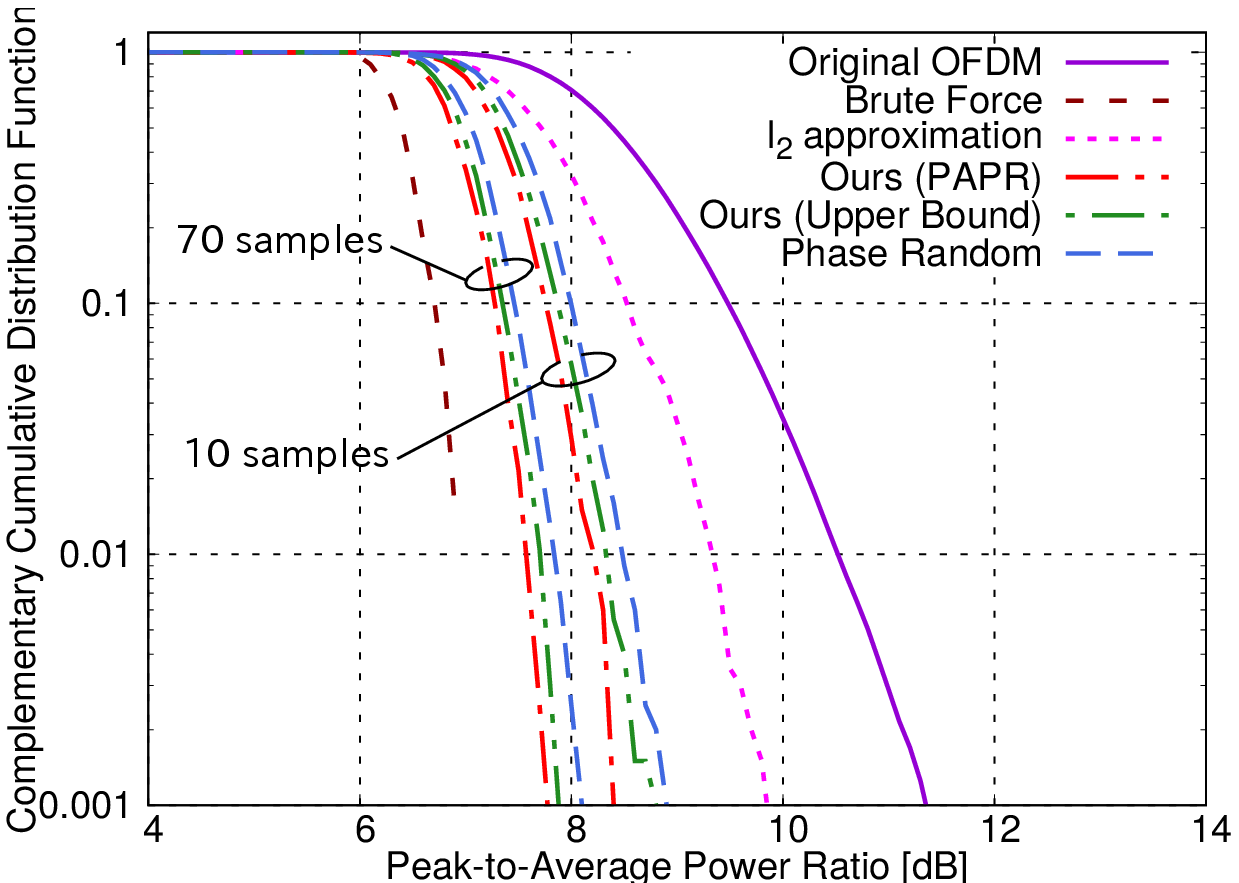}
\caption{PAPR with the parameters $(P,L)=(8,4)$}
\label{fig:P8L4}
\end{figure}

\begin{figure}[htbp]
\centering
\includegraphics[width=2.9in]{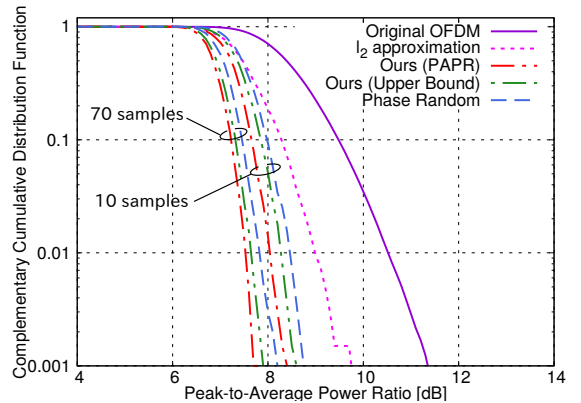}
\caption{PAPR with the parameters $(P,L)=(8,8)$}
\label{fig:P8L8}
\end{figure}

Figures \ref{fig:P16L2}, \ref{fig:P8L4} and \ref{fig:P8L8} show each PAPR curve with original OFDM systems, brute force method \cite{pts}, the $l_2$ approximation method discussed in Section IV, the randomization method discussed in Section V, the reducing upper-bound method discussed in Section VII, and the phase random method \cite{phaserandom}. In the legends, ``$l_2$ approximation'', ``Ours (PAPR)'' and ``Ours (Upper Bound)'' mean the $l_2$ approximation method, the randomization method discussed in Section V, and the reducing upper-bound method discussed in Section VII, respectively. In Fig. \ref{fig:P8L8}, the PAPR curve with the brute force method is not drawn since it is not straightforward to obtain the optimal vector due to its significantly large calculation amount. From these figures, the PAPR curve with the $l_2$ approximation method is far from one with the brute force method. This result shows that the optimal solution of the relaxed problem is far from a rank-1 matrix and it tends to have some large eigenvalues. Therefore, we conclude that the $l_2$ approximation method is not suitable for PTS techniques.

With randomization methods, there are two PAPR curves in 10 random vectors and 70 random vectors. In both numbers of random vectors, the PAPRs of our two randomization methods are lower than one of phase random techniques. As seen in Section VI, the phase random method is equivalent to our method with the identical matrix as a covariance matrix. Therefore, the performance of randomization methods can be improved when a suitable covariance matrix is chosen.

In Section VII, we have discussed the method to reduce the upper bound of PAPR. From the numerical results, PAPR with the reducing upper bound method is higher than one of the randomization method discussed in Section V. However, in a sense of solving optimization problems, the complexity with the reducing upper bound method is lower than one with the randomization method discussed in Section V. The reason is as follows.
The main point of this method is that the problem reducing upper bound of PAPR is independent of the oversampling parameter $J$. With this and the number of constraints is invariant, the complexity of the solver does not increase as $J$ increase. As seen in Eq. (\ref{eq:sample}), the sufficiently large $J$ is necessary. Then, with the randomization method discussed in Section V, the necessary number of constraints is large since $J$ is sufficiently large. Therefore, the reducing upper bound method can achieve low PAPR with low complexity.
 
\section{Conclusion}
In this paper, we have discussed how to obtain a suitable vector for partial transmit sequence techniques and have proposed two kinds of randomization methods with semidefinite relaxation techniques. Further, we have shown the relation between our methods and the phase random method. Then, in our numerical results, we have shown their PAPR curves and that our methods can achieve lower PAPR than one with the phase random method. Moreover, our numerical results have implied that randomization methods can achieve lower PAPR if a more suitable covariance matrix is obtained.

A remaining issue is to explore how to obtain a suitable covariance matrix for a randomization method. One of necessities to address this is to obtain the explicit form of a suitable covariance matrix. 
After giving such an explicit way, we expect an ideal method for obtaining low PAPR.

\appendices
\section{Proof of Convexity of Equation (\ref{eq:l2_expression2})}

In this appendix, we prove that the function defined in Eq. (\ref{eq:l2_expression2}) is convex with respect to $\mathbf{b}$
\begin{equation*}
  \|\hat{\boldsymbol{\rho}}\|_{l_2}^2 = \frac{K}{2}\sum_{k=1}^K \left\{\left(\mathbf{b}^*A^*V^*G_kVA\mathbf{b}\right)^2 + \left(\mathbf{b}^*
      A^*\hat{V}^*G_k\hat{V}A\mathbf{b}\right)^2\right\}.
\end{equation*}
First, it follows that the matrices $A^*V^*G_kVA$ and $A^*\hat{V}^*G_k\hat{V}A$ are positive semidefinite matrices since they are the Gram matrices. To prove the convexity of the above function, it is sufficient to prove that each term of the above function is convex since the sum of convex functions is convex. Therefore, we prove
\begin{equation}
  \begin{split}
  &\gamma\left(\mathbf{b}_1^*G\mathbf{b}_1\right)^2 + (1-\gamma)\left(\mathbf{b}_2^*G\mathbf{b}_2\right)^2\\
  \geq & \left(\left(\gamma\mathbf{b}_1 + (1-\gamma\mathbf{b}_2\right)^*G\left(\gamma\mathbf{b}_1 + (1-\gamma)\mathbf{b}_2\right)\right)^2,
  \end{split}
\end{equation}
where $\gamma \in [0,1]$, $\mathbf{b}_1, \mathbf{b}_2 \in \mathbb{C}^P$ and $G$ is a positive semidefinite matrix corresponding to either $A^*V^*G_nVA$ or $A^*\hat{V}^*G_n\hat{V}A$.

Let us prove the convexity. Since $x^2$ is a convex and non-decreasing function for $x \geq 0$ and $\mathbf{b}^*G\mathbf{b}$ is convex and non-negative, the following inequalities are satisfied
\begin{equation}
  \begin{split}
    & \left(\left(\gamma\mathbf{b}_1 + (1-\gamma)\mathbf{b}_2\right)^*G\left(\gamma\mathbf{b}_1 + (1-\gamma)\mathbf{b}_2\right)\right)^2\\
    \leq & \left(\gamma\mathbf{b}_1^*G\mathbf{b}_1 + (1-\gamma)\mathbf{b}_2^*G\mathbf{b}_2\right)^2.\\
    \leq & \gamma\left(\mathbf{b}_1^*G\mathbf{b}_1\right)^2 + (1-\gamma)\left(\mathbf{b}_2^*G\mathbf{b}_2\right)^2.
  \end{split}
\end{equation}
Applying the above inequalities to each term of Eq. (\ref{eq:l2_expression2}), and the sum of convex functions is convex, we have that $\|\hat{\boldsymbol{\rho}}\|^2_{l_2}$ in Eq. (\ref{eq:l2_expression2}) is convex. The same result can be obtained with the Theorem 5.1 written in \cite{cvx_analysis}.

\section{Proof of Relations in Equation (\ref{eq:upper_relation})}
In this appendix, we prove the relations written in Eq. (\ref{eq:upper_relation}) 
\begin{equation*}
  \begin{split}
   & \sum_{k=1}^K \left\{\operatorname{Tr}\left(A^*V^*G_kVAX^\star\right)^2 + \operatorname{Tr}\left(
       A^*\hat{V}^*G_k\hat{V}AX^\star\right)^2\right\}\\
   \leq & \sum_{k=1}^K \operatorname{E}\left\{\left(\boldsymbol{\xi}^*A^*V^*G_kVA\boldsymbol{\xi}\right)^2 + \left(\boldsymbol{\xi}^*
       A^*\hat{V}^*G_k\hat{V}A\boldsymbol{\xi}\right)^2\right\}\\
   \leq & 3\sum_{k=1}^K \left\{\operatorname{Tr}\left(A^*V^*G_kVAX^\star\right)^2 + \operatorname{Tr}\left(
       A^*\hat{V}^*G_k\hat{V}AX^\star\right)^2\right\}
  \end{split}
\end{equation*}
for $\boldsymbol{\xi} \sim \mathcal{CN}(\mathbf{0},X^\star)$.

A proof that the first inequality holds is given as follows. From the Cauchy-Schwarz inequality, it holds that
\begin{equation}
  \begin{split}
   & \sum_{k=1}^K \left\{\operatorname{Tr}\left(A^*V^*G_kVAX^\star\right)^2 + \operatorname{Tr}\left(
       A^*\hat{V}^*G_k\hat{V}AX^\star\right)^2\right\}\\
   \leq & \sum_{k=1}^K \operatorname{E}\left\{\left(\boldsymbol{\xi}^*A^*V^*G_kVA\boldsymbol{\xi}\right)^2 + \left(\boldsymbol{\xi}^*
       A^*\hat{V}^*G_k\hat{V}A\boldsymbol{\xi}\right)^2\right\}.
  \end{split} 
\end{equation}

Then, a proof that the last inequality holds is given as follows.
Similar to Appendix A, it is sufficient to prove
\begin{equation}
 \operatorname{E}\left\{\left(\boldsymbol{\xi}^* G \boldsymbol{\xi}\right)^2\right\} 
\leq 3 \operatorname{Tr}\left(G X^\star\right)^2,
\label{eq:proof_obj}
\end{equation}
where $G$ is a Hermitian and positive semidefinite matrix and $\boldsymbol{\xi} \sim \mathcal{CN}(\mathbf{0},X^\star)$. In \cite{graphical}, it has been shown that $\mathcal{T}(\boldsymbol{z}) \sim \mathcal{N}\left(\mathcal{T}(\boldsymbol{\mu}),\frac{1}{2}\mathcal{T}(\Sigma))\right)$ if $\boldsymbol{z} \sim \mathcal{CN}(\boldsymbol{\mu},\Sigma)$. Note that the matrices $\mathcal{T}(X^\star)$ and $\mathcal{T}(G)$ are symmetric and positive semidefinite since $G$ and $X^\star$ are Hermitian and positive semidefinite \cite{telatar}.
From this result, it follows that $\mathcal{T}(\boldsymbol{\xi}) \sim \mathcal{N}\left(\mathcal{T}(\mathbf{0}),\frac{1}{2}\mathcal{T}(X^\star)\right)$. With this and discussions in \cite{telatar}, the left hand side of Eq. (\ref{eq:proof_obj}) is rewritten as
\begin{equation}
\begin{split}
 &\operatorname{E}\left\{\left(\boldsymbol{\xi}^* G \boldsymbol{\xi}\right)^2\right\}\\
=  & \operatorname{E}\left\{\left(\mathcal{T}(\boldsymbol{\xi})^\top \mathcal{T}(G) \mathcal{T}(\boldsymbol{\xi})\right)^2\right\}\\
= & \sum_{i,j,k,l} \hat{g}_{i,j}\hat{g}_{k,l}\operatorname{E}\{\hat{\xi}_i\hat{\xi}_j\hat{\xi}_k\hat{\xi}_l\},
\end{split}
\end{equation}
where $\hat{g}_{i,j}$ and $\hat{\xi}_k$ are the $(i,j)$-th element of $\mathcal{T}(G)$ and $k$-th element of $\mathcal{T}(\boldsymbol{\xi})$, respectively. In \cite{multivariate}, for $\boldsymbol{x} \sim \mathcal{N}(\boldsymbol{\mu},\Sigma)$, the forth moment about the mean has been derived as
\begin{equation}
 \begin{split}
& \operatorname{E}\{(x_i - \mu_i)(x_j - \mu_j)(x_k - \mu_k)(x_l - \mu_l)\}\\
= & \sigma_{i,j}\sigma_{k,l} + \sigma_{i,k}\sigma_{j,l} + \sigma_{i,l}\sigma_{j,k},
  \end{split}
\label{eq:4thmoment}
\end{equation}
where $x_i$, $\mu_i$ and $\sigma_{i,j}$ are the $i$-th element of $\boldsymbol{x}$, the $i$-th element of $\boldsymbol{\mu}$ and the $(i,j)$-th element of the real valued-covariance matrix $\Sigma$, respectively. With Eq. (\ref{eq:4thmoment}), Eq. (\ref{eq:proof_obj}) is rewritten as
\begin{equation}
\begin{split}
 &\operatorname{E}\left\{\left(\boldsymbol{\xi}^* G \boldsymbol{\xi}\right)^2\right\}\\
= & \frac{1}{4}\left\{\sum_{i,j,k,l} \hat{g}_{i,j}\hat{g}_{k,l}(\hat{x^\star}_{i,j}\hat{x^\star}_{k,l} + \hat{x^\star}_{i,k}\hat{x^\star}_{j,l} + \hat{x^\star}_{i,l}\hat{x^\star}_{j,k})\right\}\\
= & \frac{1}{4}\left\{\operatorname{Tr}(\mathcal{T}(G)\mathcal{T}(X^\star))^2 + 2\operatorname{Tr}(\mathcal{T}(G)\mathcal{T}(X^\star)\mathcal{T}(G)\mathcal{T}(X^\star))\right\},
\end{split}
\label{eq:4thmoment_expand}
\end{equation}
where $\hat{x^\star}_{i,j}$ is the $(i,j)$-th element of $\mathcal{T}(X^\star)$. In deriving the above second equality, we have used the property that the matrices $\mathcal{T}(G)$ and $\mathcal{T}(X^\star)$ are symmetric. Let $V$ be a matrix such that $VV^\top = \mathcal{T}(G)$, where such a $V$ can be found since $\mathcal{T}(G)$ is a positive semidefinite. With this decomposition, the relations
\begin{equation}
 \begin{split}
& \operatorname{Tr}(\mathcal{T}(G)\mathcal{T}(X^\star)\mathcal{T}(G)\mathcal{T}(X^\star))\\
= & \operatorname{Tr}(VV^\top\mathcal{T}(X^\star)VV^\top\mathcal{T}(X^\star))\\
= & \operatorname{Tr}(V^\top\mathcal{T}(X^\star)V \cdot V^\top\mathcal{T}(X^\star)V)\\
\leq & \operatorname{Tr}(V^\top\mathcal{T}(X^\star)V)^2\\ 
= & \operatorname{Tr}(\mathcal{T}(X^\star)\mathcal{T}(G))^2
\end{split}
\label{eq:4thmoment_right}
\end{equation} 
are obtained.
In the above relations, we have used the properties that the matrix $V^\top\mathcal{T}(X^\star)V$ is positive semidefinite and $\operatorname{Tr}(XX) \leq \operatorname{Tr}(X)^2$ for any positive semidefinite matrix $X$. In \cite{maxcut}, it has been shown that $\operatorname{Tr}(\mathcal{T}(X)\mathcal{T}(Y)) = 2\operatorname{Tr}(XY)$ for positive semidefinite matrices $X$ and $Y$. Combining this result and Eqs. (\ref{eq:4thmoment_expand}) (\ref{eq:4thmoment_right}), we arrive at the relation
\begin{equation}
\operatorname{E}\left\{\left(\boldsymbol{\xi}^* G \boldsymbol{\xi}\right)^2\right\} \leq  3 \operatorname{Tr}\left(G X^\star\right)^2.
\end{equation}
This is the desired result.

We have proven Eq. (\ref{eq:proof_obj}) for general $L$. For $L=2$ and $L=4$, we have the same expressions of Eq. (\ref{eq:proof_obj}).


%


\section*{Acknowledgment}
The authors would like to thank Dr. Shin-itiro Goto for his advise.


\ifCLASSOPTIONcaptionsoff
  \newpage
\fi

\end{document}